\documentclass[journal]{IEEEtran}
\IEEEoverridecommandlockouts
\usepackage{amsmath,amsfonts}
\usepackage{mathtools}
\usepackage{amsthm}
\usepackage{array}
\usepackage{textcomp}
\usepackage{stfloats}
\usepackage{url}
\usepackage{microtype}
\usepackage{verbatim}
\usepackage{graphicx}
\usepackage{cite}
\usepackage[colorlinks=true, linkcolor=blue, citecolor=blue, urlcolor=blue]{hyperref}
\usepackage{enumitem}

\usepackage[linesnumbered,lined,algoruled,commentsnumbered]{algorithm2e}

\usepackage{caption}
\usepackage{subcaption}
\usepackage{multirow}
\usepackage{booktabs}
\usepackage[table]{xcolor}
\usepackage{stfloats}
\usepackage{lipsum} 
\usepackage[font=small,skip=0pt]{caption}
\usepackage[export]{adjustbox}
\usepackage[inkscapelatex=false]{svg}

\SetCommentSty{mycommfont}
\theoremstyle{definition}

\usepackage{units}
\newcommand{\nth}[1]{{#1}^{\text{th}}}

\newcommand{\abs}[1]{\left|{#1}\right|}

\newcommand{\RD}[0]{r_{\mathrm{\scalebox{0.5} {RD} }}}

\newcommand{\rf}[0]{r_{\mathrm{\scalebox{0.5} {0}}}}
\newcommand{\au}[0]{\theta_{\mathrm{\scalebox{0.5} {0}}}}

\usepackage{acronym}
\input{acronym}

\usepackage{titlesec}
\titlespacing*{\section}{0pt}{1pt}{1pt} 
\titlespacing*{\subsection}{0pt}{1pt}{.5pt}
\titlespacing*{\subsubsection}{0pt}{.75pt}{.5pt}

\begin{document}
\title{Sparse Array Design for Near-Field MU-MIMO: Reconfigurable Array Thinning Approach

\author{Ahmed Hussain~\IEEEmembership{Graduate Student Member, IEEE}, Asmaa Abdallah,~\IEEEmembership{Senior Member, IEEE}, Abdulkadir Celik,~\IEEEmembership{Senior Member, IEEE}, Emil Bj\"ornson,~\IEEEmembership{Fellow, IEEE} and Ahmed M. Eltawil,~\IEEEmembership{Senior Member, IEEE}

\thanks{Ahmed Hussain, Asmaa Abdallah, and Ahmed M. Eltawil are with Computer, Electrical, and Mathematical Sciences and Engineering (CEMSE) Division, King Abdullah University of Science and Technology (KAUST), Thuwal, 23955-6900, KSA. Abdulkadir Celik is with School of Electronics and Computer Science, University of Southampton, SO17 1BJ UK. Emil Bj\"ornson is with the School of Electrical Engineering and Computer Science, KTH Royal Institute of Technology, 100 44 Stockholm, Sweden. The work of E.~Bj\"ornson was supported by the Grant 2022-04222 from the Swedish Research Council.}
}
}

\maketitle

\begin{abstract}
Future wireless networks, deploying thousands of antenna elements, may operate in the radiative near-field (NF), enabling spatial multiplexing across both angle and range domains. Sparse arrays have the potential to achieve comparable performance with fewer antenna elements. However, fixed sparse array designs are generally suboptimal under dynamic user distributions, while movable antenna architectures rely on mechanically reconfigurable elements, introducing latency and increased hardware complexity. To address these limitations, we propose a reconfigurable array thinning approach that selectively activates a subset of antennas to form a flexible sparse array design without physical repositioning. We first analyze grating lobes for uniform sparse arrays in the angle and range domains, showing their absence along the range dimension. Based on the analysis, we develop two particle swarm optimization-based strategies: a grating-lobe-based thinned array (GTA) for grating-lobe suppression and a sum-rate-based thinned array (STA) for multiuser sum-rate maximization. Simulation results demonstrate that GTA outperforms conventional uniform sparse arrays, while STA achieves performance comparable to movable antennas, thereby offering a practical and efficient array deployment strategy without the associated mechanical complexity.
\end{abstract}

\section{Introduction}\label{Sec-I}
\IEEEPARstart{F}{uture} wireless networks are expected to deploy increasingly large antenna arrays, thereby extending communication into the radiating \ac{NF} regime \cite{11095387}. Unlike the \ac{FF}, where \acp{UE} are multiplexed solely in the angular domain, spherical wave propagation in the \ac{NF} enables finite-depth beamforming that resolves \acp{UE} jointly in angle and range. This additional spatial dimension enhances spatial multiplexing gains \cite{10934779}. However, realizing large-aperture arrays with half-wavelength spacing requires thousands of antenna elements, leading to a significant increase in hardware cost, power consumption, and computational complexity. Achieving high spatial multiplexing gain in multi-user \ac{MIMO} systems necessitates strong orthogonality among the \ac{UE} channel vectors to suppress inter-user interference. Sparse arrays offer a potential solution, especially in low-scattering environments, by leveraging larger inter-element spacings to reduce spatial correlation and generate more diverse channels than half-wavelength arrays \cite{11030818}. Nevertheless, uniform sparse arrays suffer from grating lobes, which cause strong interference by illuminating unintended directions. To overcome this limitation, non-uniform array geometries, such as coprime arrays~\cite{zhou2024sparse} and array-position optimization techniques~\cite{10563980,bjornson2026mag} have been investigated. More recently, movable antenna (MA) architectures have been proposed, wherein antenna positions are adjusted to match the instantaneous user distribution~\cite{zhu2025tutorial}. Despite their potential, MA-based designs rely on physical repositioning, which introduces latency overheads and system-level complexity. A key limitation of existing sparse array solutions is that they are either static, optimized for a specific user distribution and therefore suboptimal under different channel conditions, or mechanical, as in MA architectures that require continuous physical movement and high-precision hardware. Both approaches are thus challenging to implement in practical deployments.

To address these challenges, we propose a reconfigurable array thinning framework in which the full dense array remains physically fixed, while only a subset of antennas is activated to meet a desired performance objective. It is important to distinguish conventional antenna selection approaches in communication systems from array thinning-based sparse array design. Existing antenna selection methods rely on \ac{CSI} to identify a subset of antennas that contributes most significantly to system performance, using metrics such as singular values of the channel matrix, channel norms, spatial correlation, or minimum \ac{SNR} \cite{11137408}. In this context, antenna selection is not explicitly formulated as a sparse array design problem. In contrast, array thinning aims to design structured sparse arrays, where the thinning operation can be performed either offline before deployment based on statistical \ac{CSI} \cite{bjornson2026mag}, or online by adapting to user location information obtained, for example, through beam training. To the best of the authors' knowledge, this is the first work that considers array thinning for \ac{NF} communication. We investigate three fundamental research gaps: First, although it is well known that exceeding half a wavelength yields grating lobes in the angular domain, it remains unclear whether similar phenomena also arise in the range domain. Second, we examine the achievable multiuser sum-rate when a thinned array is pre-optimized for grating-lobe suppression, and assess the performance gain compared to uniform sparse arrays. Third, we investigate how to design dynamic thinned arrays that maximize the multiuser sum-rate. To address these questions, we first analyze grating lobes in the \ac{NF} across both the angle and range domains. Building on the insights drawn from the analysis of grating lobes, we propose two types of sparse array designs using a \ac{PSO}-based optimization framework. The first is a \textit{pre-optimized} design, termed \ac{GTA}, which aims to suppress grating lobes in the \ac{NF}. The second design, \ac{STA}, is \textit{dynamically} optimized to directly maximize the multi-user sum-rate. In addition, inspired by the pre-optimized sparse \ac{MULA} in~\cite{11030818}, where antenna positions are optimized based on statistical \ac{CSI}, we extend this concept to the \ac{NF} regime by performing array thinning instead of position optimization, which we refer to as \ac{PTA}. We benchmark the proposed designs against other sparse arrays, including \ac{SULA} and \ac{MULA}. 


\section{System Model}\label{Sec-II}
We consider a \ac{BS} equipped with a \ac{FULA} of $N$ antennas with inter-element spacing $d=\tfrac{\lambda}{2}$, resulting in an aperture length $D=(N-1)\tfrac{\lambda}{2}$. During operation, the \ac{BS} activates only $N_{\mathrm{T}}$ antennas, forming a thinned array characterized by the thinning ratio $\operatorname{TR}=\tfrac{N_{\mathrm{T}}}{N}$. In a downlink free-space \ac{LoS} scenario, the \ac{BS} simultaneously serves $K$ single-antenna \acp{UE}, where each data symbol $s_k$ is precoded using the beamforming vector $\mathbf{w}_k \in \mathbb{C}^{N}$ and transmitted from the \ac{BS}. The received signal at the $\nth{k}$ \ac{UE} is expressed as
\begin{equation}
y_{k} = \mathbf{w}_{k}^\mathsf{H} \mathbf{h}_{k} s_{k} + \sum_{\substack{j=1, \ j \neq k}}^{K} \mathbf{w}_{k}^\mathsf{H} \mathbf{h}_{j} s_{j} + z_{k},
\label{eqn_A1}
\end{equation}
where $z_{k}$ represents additive circularly symmetric complex Gaussian noise with variance $\sigma^2$. The channel vector $\mathbf{h}_{k} \in \mathbb{C}^{N}$ between the \ac{BS} and the $\nth{k}$ \ac{UE}, is given by
\begin{equation}
\mathbf{h}_{k}
=\sqrt{\beta_{k}}e^{-j\tfrac{2\pi}{\lambda}r_k}\,\big( \mathbf{b} \odot \mathbf{a}(\theta_{k}, r_{k}) \big),
\qquad
\beta_{k}=\frac{\lambda^{2}}{(4\pi)^{2} r_{k}^{2}},
\label{eqn_IIA_2}
\end{equation}
where $\odot$ denotes the element-wise (Hadamard) product. The coefficient $\beta_{k}$ captures the path loss, and $\mathbf{a}(\theta_{k}, r_{k}) \in \mathbb{C}^{N}$ is the \ac{NF} array response vector corresponding to azimuth angle $\theta_{k}$ and range $r_{k}$. The vector $\mathbf{b} \in \{0,1\}^{N}$ represents the binary thinning weights. An entry $b_{n} = 1$ indicates that the $\nth{n}$ antenna element is active, whereas $b_{n} = 0$ denotes a deactivated (thinned) element. In practice, this can be implemented using a dynamic sub-array architecture, in which an RF switch network allows each antenna element to be connected to any \ac{RF} chain, while inactive elements are terminated to ground. A more power-efficient alternative is a fixed sub-array-based architecture, where each \ac{RF} chain is restricted to a predefined subset of antenna elements through a switch network. These architectures involve a fundamental trade-off between spectral efficiency and power efficiency. A detailed analysis of this trade-off is left for future work. The normalized \ac{NF} array response vector for the $\nth{n}$ antenna is given as \cite{10988573} 
\begin{equation} 
{a}_n (\theta,r) \approx \tfrac{1}{\sqrt{N}}e^{-j\tfrac{2\pi}{\lambda}\{nd\sin(\theta)- \frac{1}{2r}n^2d^2\cos^2(\theta)\}},
\label{eqn-A4}
\end{equation}
which is valid when the \ac{UE} range exceeds twice the aperture length, i.e., $r > 2D$. To suppress interference in \eqref{eqn_A1}, we employ a regularized zero-forcing precoder. The resulting achievable sum-rate is 
\begin{equation}\small
\mathcal{R}_{\mathrm{sum}}=\sum_{k=1}^K\log _{2}\left(1+\operatorname{\Gamma}_{k}\right),
\label{eqn_IIB_1}
\end{equation}
\normalsize
where $\operatorname{\Gamma}_{k}$ denotes the signal-to-interference-plus-noise ratio of the $\nth{k}$ \ac{UE} and is given by,
\begin{equation}\small
\operatorname{\Gamma}_{k}=\frac{\left|\mathbf{w}_{k}^\mathsf{H}\mathbf{h}_{k} \right|^{2}}{\sigma^2+ \sum_{j=1, j \neq k}^{K}\left| \mathbf{w}_{j}^\mathsf{H}\mathbf{h}_{k} \right|^{2}}. 
\label{eqn_IIB_2}
\end{equation}\normalsize
For a given \ac{UE} distribution, the sum-rate in \eqref{eqn_IIB_1} can be optimized by adjusting antenna positions. However, real-time repositioning incurs high complexity, cost, and latency, as the optimization must be performed whenever the geometrical parameters of the channel change. To overcome this, we propose an array thinning strategy that activates only a subset of antennas to achieve performance comparable to the \ac{MULA}. We formulate the antenna-activation task as an optimization over the activation vector $\mathbf{b} = [b_{1},\dots,b_{N}]^\mathsf{T}$. Let $f(\mathbf{b})$ denote a generic objective function (e.g., sum-rate). The antenna-selection problem is expressed as
\begin{equation}
\begin{aligned}
\max\limits_{\mathbf{b}} \quad & f(\mathbf{b}) \ \ 
\text{s.t.} \
& \sum_{n=1}^{N} b_{n} = N_{\mathrm{T}},
\end{aligned}
\label{eq:gen_opt_problem}
\end{equation}
The constraint $\sum_{n=1}^{N}b_{n} = N_{\mathrm{T}}$ enforces the thinning ratio, ensuring that exactly $N_{\mathrm{T}}$ antennas remain active. While the overarching goal remains sum-rate maximization, we employ two alternative objective functions $f(\mathbf{b})$ which are detailed in Section~\ref{Sec-IV}.
\section{Grating Lobes in the Near-field} \label{Sec-III}
In this section, we investigate the occurrence of grating lobes in both the angular and range dimensions.
\subsection{Grating Lobes in the Angle Domain} \label{Sec:AngleGrating}
Consider an \ac{NF} beam focused at the location $(\au,\rf)$. The corresponding beam pattern in the angle domain is obtained as
\begin{align}\small
\mathcal{G}(\theta)
&= \left| \mathbf{a}^{\mathsf{H}}(\au,\rf) \, \mathbf{a}(\theta,r) \right|^2 \nonumber\\
&= \left| \frac{1}{N} \sum_{n=0}^{N-1} e^{\,j\tfrac{2\pi}{\lambda} n d (\sin\theta - \sin\au)} e^{-j\tfrac{2\pi}{\lambda} \tfrac{n^2 d^2}{2}\!\left( \tfrac{\cos^2\theta}{r} - \tfrac{\cos^2\au}{\rf} \right)} \right|^2 \nonumber\\
&\stackrel{(a)}
{\approx}
\left| \frac{1}{N} \sum_{n=0}^{N-1} 
e^{\,j\tfrac{2\pi}{\lambda} n d (\sin\theta - \sin\au)} \right|^2,
\label{eq:arrayfactor}
\end{align}
\normalsize
where approximation ~$(a)$ follows from the distance-ring condition $\frac{\cos^2 (\theta)}{r} = \frac{\cos^2 (\au)}{\rf}$. This condition defines a set of range–angle pairs along which the beamwidth remains nearly constant. The simplified expression in~\eqref{eq:arrayfactor} represents the array factor in the angle domain. We present Property~1 to review grating lobes in the angle domain and then use it to analyze grating lobes in the range domain.
\newtheorem{property}{Property}
\begin{property}[Periodicity Condition of the Array Factor]
Consider the discrete sequence $e^{j\phi_n(x)}$, where $x \in \{\theta, r\}$ and $\phi_n(x)$ denotes the phase contribution of the $\nth{n}$ antenna element. The sequence $e^{j\phi_n(x)}$ is periodic if and only if there exists a constant increment $\Delta x$ such that \cite{hansen2009phased}
\begin{equation}
    \phi_n(x) - \phi_n(x_0) = 2\pi q,
    \qquad q \in \mathbb{Z},
\label{eq:discrete_periodicity_property}
\end{equation}
where $x_0$ denotes the focused angle/range, and $x = x_0 + \Delta x$ is the observation angle/range at which the presence of a grating lobe is evaluated.
\end{property}
 Applying \eqref{eq:discrete_periodicity_property} to the array factor expression in \eqref{eq:arrayfactor}, the grating lobes occur when the phase shift between adjacent elements equals an integer multiple of $2\pi$ \cite{hansen2009phased}
\begin{equation}
2\pi \frac{d}{\lambda}(\sin\theta - \sin\au) = 2\pi q,
\quad q \in \mathbb{Z}.
\label{eq:gratingcond1}
\end{equation}
Solving for $\theta$ yields the following grating lobe condition:
\begin{equation}
\sin\theta_q = \sin\au + q\frac{\lambda}{d},
\quad q = \pm1, \pm2, \ldots
\label{eq:gratingcond2}
\end{equation}
Grating lobes appear at angles $\theta_q$ when the mainlobe is focused at $\au\in[-90^\circ,90^\circ]$, provided that $|\sin\theta_q| = |\sin\au + q\frac{\lambda}{d} |\le 1$. For a \ac{ULA}, the visible angular region is given by $\theta \in [-90^\circ,90^\circ]$, so no grating lobes occur within this region when $d \le \lambda/2$. In contrast, for $d > \lambda/2$, additional lobes appear at angular locations determined by~\eqref{eq:gratingcond2}. For example, when $d = 2\lambda$ and $\au = 0^\circ$, \eqref{eq:gratingcond2} reduces to $\sin\theta_q = \frac{q}{2}$, which satisfies $|\sin\theta_q|\leq 1$ for $q \in \{-1, 1\}$. The corresponding grating lobe angles are $\theta_q \in \{-30^\circ,30^\circ\}$.

\subsection{Absence of Grating Lobes in the Range Domain} \label{Sec 3-B}
\begin{figure}[t]
\centering
\begin{minipage}{1\columnwidth}  
    \centering
    \includegraphics[width=\linewidth]{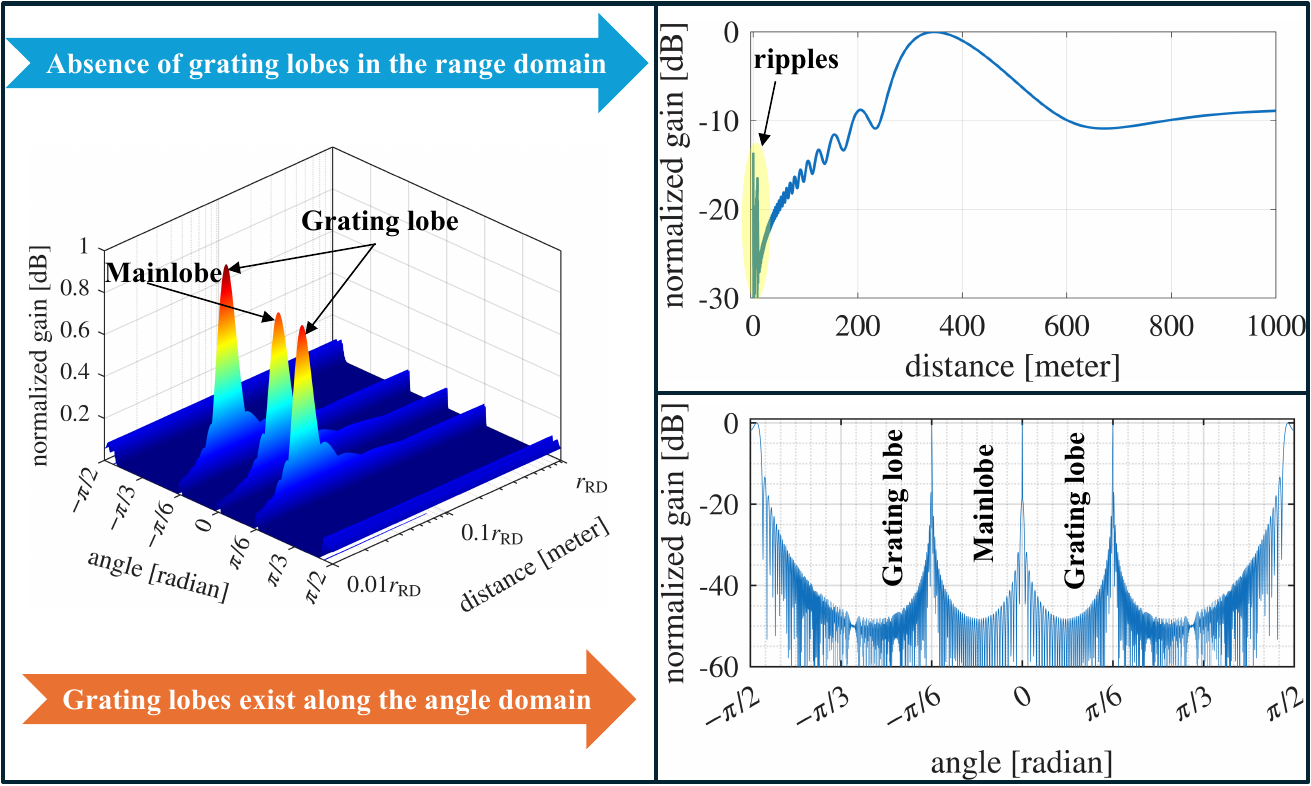}  
    \captionsetup{skip=2pt}  
    \caption{Beam pattern in angle and range domain: grating lobes appear only in the angular domain. Here we set $f_c = \unit[15]{GHz}$, $N = 256$, $d=2\lambda$, $\rf = \unit[346]{m}$ and $\RD=\unit[10.3]{km}$.}
    \label{fig:grating_near}
\end{minipage}
\vspace{-20pt}  
\end{figure}

The beam pattern in the range domain is obtained as the inner product of \ac{NF} array response vectors given in \eqref{eqn-A4}, pointing to the same angle $\theta$ but different distances $r$ and $\rf$ \cite{10988573}
\begin{align}\small
\mathcal{G}(\theta,r) &= \left| \mathbf{a}^\mathsf{H} (\theta, \rf) \, \mathbf{a} (\theta, r) \right|^2, \label{eqn-B6-15}\\
\overset{(\mathrm{a})}{=} \ & \left| \frac{1}{N} \sum_{n=0}^{N-1} e^{ -j\tfrac{2\pi}{\lambda} n^2 d^2 \cos^2(\theta) r_\mathrm{eff} } \right|^2, 
\label{eqn-B6-17}
\end{align}\normalsize
where $r_\mathrm{eff} = \abs{\frac{r - \rf}{2r\rf}} $ in (a). To assess the possibility of grating lobes in the range domain, the phase in~\eqref{eqn-B6-17} must satisfy the periodicity condition in~\eqref{eq:discrete_periodicity_property}:
\begin{align}
\tfrac{2\pi}{\lambda} d^2 \cos^2(\theta)\, r_\mathrm{eff} = 2\pi q, 
\qquad \ q \in \mathbb{Z}, 
\label{eqn-B17}
\end{align}
Solving for $r$ yields the following expression:
\begin{align}
r_q = \frac{\rf d^2 \cos^2(\theta)}{d^2 \cos^2(\theta) + 2 q \rf \lambda}.
\label{eqn-B18}
\end{align}
However, unlike the angular domain, where the phase varies linearly with element index $n$, the phase in \eqref{eqn-B6-17} varies quadratically. As a result, at the distances $r_q$ given by \eqref{eqn-B18}, the individual element phases do not re-align coherently across the array, and the summation over $n$ does not produce a secondary mainlobe; instead, only small ripples are formed. Furthermore, the resulting $r_q$ values are either negative (for $q<0$) or extremely small (for $q>1$), rendering them physically impractical. Therefore, unlike the angular domain, grating lobes do not appear along the range dimension. To exemplify this with a numerical example, we consider a $256$ element \ac{SULA} with inter-element spacing $d = 2\lambda$ and an \ac{NF} beam focused at $(\au = 0^\circ, \rf = \RD/30= \unit[346]{m})$, where $\RD = \tfrac{2D^{2}}{\lambda}$ denotes the Rayleigh distance. As shown in Fig.~\ref{fig:grating_near}, we plot the resulting \ac{2D} beam pattern along with the corresponding \ac{1D} cuts in angle and range. In the \ac{2D} plot, two additional grating lobes emerge at $\theta = \pm \tfrac{\pi}{6}$, consistent with the angular positions predicted in the last subsection. In contrast, the range-domain response exhibits no grating lobes. This distinction is further clarified by the \ac{1D} angular and range patterns, which highlight the presence of grating lobes solely in the angular domain. Furthermore, small ripples are observed at short ranges in the 1D range-domain response. Specifically, the dominant ripple attains a level of $\unit[-13]{dB}$ at a distance of $\unit[0.01]{m}$.

\section{Proposed Optimization Methods} \label{Sec-IV}
In this section, we explain the two proposed array-thinning strategies for optimizing the sum-rate in \eqref{eqn_IIB_1}. The first method, \ac{GTA}, suppresses grating lobes, while the second method, \ac{STA}, directly maximizes the multiuser sum-rate.
\subsection{PSO-based Array Thinning for Grating-Lobe Suppression}
Grating lobes can be mitigated by disrupting the periodic structure of the \ac{SULA}. Based on the results in Section~\ref{Sec-III}, grating lobes are required to be suppressed only in the angular domain. Furthermore, the beam pattern in the angular domain remains invariant between the \ac{NF} and \ac{FF}. Similarly, as indicated by \eqref{eq:gratingcond2}, the locations of grating lobes do not depend on the range. Therefore, we consider the weighted angle-domain \ac{FF} beam pattern of a \ac{ULA} with $N$ antennas, given by
\begin{equation}
\small
   \mathcal{G}(\mathbf{b},\theta) 
   = \left| \frac{1}{N} 
   \sum_{n=0}^{N-1} 
   b_n\ \odot e^{\,j\tfrac{2\pi}{\lambda} n d (\sin\theta - \sin\au)} \right|^2,
\end{equation}
\normalsize
 The thinning problem aims to determine an optimal binary vector $\mathbf{b}$ that minimizes the \ac{PSL} within a specified angular coverage region $\Theta_\mathrm{cov}$. Since the grating-lobe behavior varies with the angle $\au$, the \ac{PSL} is defined as a function of $\au$ as
 \begingroup
\setlength{\abovedisplayskip}{.7pt}   
\setlength{\belowdisplayskip}{.7pt}   
\begin{equation}
\small
    \mathrm{PSLL}(\mathbf{b}, \au) 
    = 10 \log_{10} \left(
        \frac{
            \displaystyle \max_{\theta \in \mathcal{S}} 
            \mathcal{G}(\mathbf{b},\theta)
        }{
            \mathcal{G}(\mathbf{b},\au)
        }
    \right),
    \label{eqn:PSLL}
\end{equation}
\endgroup
\normalsize
where $\mathcal{S}$ denotes the sidelobe region excluding the mainlobe around $\au$. As the steering angle $\au$ increases, additional grating lobes may appear. For a given antenna spacing $d$, more grating lobes appear as the angle is steered towards the endfire direction. Hence, we aim to design a thinning pattern that minimizes the \ac{PSL} over the maximum steering angle within the coverage interval $\Theta_\mathrm{cov}$. Accordingly, the optimization problem \eqref{eq:gen_opt_problem}, where $f(\mathbf{b})=\mathrm{PSLL}(\mathbf{b},\au)$, is formulated as
\begin{align}
    \min_{\mathbf{b} } \quad &
        \max_{\au \in \Theta_\mathrm{cov}} \; \mathrm{PSLL}(\mathbf{b},\au),
        \label{opt:psll-min} \\[2mm]
    \text{s.t.} \quad &
        \mathrm{PSLL}(\mathbf{b},\au) \le \tau_\mathrm{PSLL},
        \label{opt:psll-threshold} \\[1mm]
    & \sum_{n=1}^{N} b_n = N_\mathrm{T},
        \label{opt:active-count} \\[1mm]
    & b_n = 1,\qquad n \in \mathcal{F},
        \label{opt:mandatory-elements}
\end{align}
where $\tau_\mathrm{PSLL}$ denotes the maximum allowable sidelobe level. The constraint in \eqref{opt:active-count} enforces a fixed number of active antennas, while \eqref{opt:mandatory-elements} ensures that a predefined set $\mathcal{F}$ of mandatory active elements (e.g., the two edge elements) is always preserved to maintain the maximum aperture. A direct binary optimization over the thinning vector is computationally intractable due to the combinatorial search space. To address this, we leverage \ac{PSO} that optimizes the objective function by iteratively evaluating $P$ candidate solutions. In \ac{PSO}, each particle $p$ represents a potential solution. More specifically, we adopt a \ac{PSO}-based continuous relaxation, where each particle represents a continuous priority vector $\mathbf{x}^{(p)} \in [0,1]^{N_v}$. Here, $N_v = N - |\mathcal{F}|$ denotes the number of variable (non-fixed) antenna positions. This vector is subsequently mapped to the binary thinning vector $\mathbf{b}^{(p)}$. The complete procedure is summarized in Algorithm~\ref{alg:PSO_thinning}. The initialization (lines 3–5) assigns each particle a random position $\mathbf{x}^{(p)}(0)$ and velocity $\mathbf{v}^{(p)}(0)$, constructs the corresponding binary vector $\mathbf{b}^{(p)}(0)$, and evaluates the initial cost $f^{(p)}(0)= f(\mathbf{b}^{(p)}(0))=\mathrm{PSLL}(\mathbf{b}^{(p)}(0),\au) $. Each particle stores its personal best $\mathbf{P}_{\mathrm{best}}^{(p)}$ and the globally best particle determines $\mathbf{G}_{\mathrm{best}}$. At iteration $t$, the velocity of the particle $p$ is updated (line 8) according to
\begin{align}
\mathbf{v}^{(p)}(t{+}1)
=\,& \omega\,\mathbf{v}^{(p)}(t)
+ c_1 u_1\!\left(\mathbf{P}_{\mathrm{best}}^{(p)}-\mathbf{x}^{(p)}(t)\right) 
+  \notag\\ 
&c_2 u_2\!\left(\mathbf{G}_{\mathrm{best}}-\mathbf{x}^{(p)}(t)\right),
\label{eqn_vel_update}
\end{align}
where $\omega$ is the inertia weight, $c_1$ and $c_2$ are acceleration coefficients, and $u_1,u_2 \sim \mathcal{U}(0,1)$ are random scalars. The updated velocity yields a new position (line 9):
\begin{equation}
\mathbf{x}^{(p)}(t{+}1)
= \mathbf{x}^{(p)}(t) + \mathbf{v}^{(p)}(t{+}1).
\label{eqn_pos_update}
\end{equation}
The entries of $\mathbf{x}^{(p)}(t{+}1)$ are clipped to the interval $[0,1]$ (line 10). In line 11, a new thinning vector $\mathbf{b}^{(p)}(t{+}1)$ is constructed by activating the antenna indices corresponding to the $\text{Top}_{N_{\mathrm{T}}-|\mathcal{F}|}$ entries of $\mathbf{x}^{(p)}(t{+}1)$ and appending the fixed set $\mathcal{F}$. The corresponding objective function $f^{(p)}(t{+}1)$ is evaluated using~\eqref{eqn:PSLL}. Each particle updates its personal best $\mathbf{P}_{\mathrm{best}}^{(p)}$ whenever $f^{(p)}(t{+}1)$ improves upon its previous value, and the global best $\mathbf{G}_{\mathrm{best}}$ is replaced whenever a particle attains the lowest cost across the swarm. This process repeats for $n_{\mathrm{PSO}}$ iterations, after which the optimal thinning vector $\mathbf{b}_{\mathrm{opt}}$ is obtained by mapping $\mathbf{G}_{\mathrm{best}}$ to its binary representation.
 
\begin{algorithm}[t]\footnotesize
\caption{PSO for Array Thinning }%
\label{alg:PSO_thinning}
\DontPrintSemicolon
\textbf{Input:} Number of particles $P$, iterations $n_{\mathrm{PSO}}$,  
variable indices $N_v$, fixed set $\mathcal{F}$, active antennas $N_{\mathrm{T}}$,
PSO parameters $(\omega, c_1, c_2)$ \\
\textbf{Output:} Optimal thinning vector $\mathbf{b}_{\mathrm{opt}}$

\textbf{Initialization:} 
$\mathbf{x}^{(p)}(0) \in [0,1]^{N_v},\; \mathbf{v}^{(p)}(0) \in \mathbb{R}^{N_v}$,\\
$\mathbf{b}^{(p)}(0) = \text{Top}_{N_{\mathrm{T}}-|\mathcal{F}|}(\mathbf{x}^{(p)}(0)) \cup \mathcal{F},\;
f^{(p)}(0) = f(\mathbf{b}^{(p)}(0))$,\\
$\mathbf{P}_{\mathrm{best}}^{(p)} = \mathbf{x}^{(p)}(0),\;
\mathbf{G}_{\mathrm{best}} = \mathbf{x}^{(p^\star)}(0),\;
p^\star = \arg\min_{p} f^{(p)}(0)$

\For{$t = 0$ \KwTo $n_{\mathrm{PSO}}-1$}{
    \For{$p = 1$ \KwTo $P$}{
        $\mathbf{v}^{(p)}(t{+}1) \leftarrow $  \tcp{ Refer Eq. \eqref{eqn_vel_update}} 
        $\mathbf{x}^{(p)}(t{+}1) \leftarrow $ \tcp{ Refer Eq. \eqref{eqn_pos_update}}
        $\mathbf{x}^{(p)}(t{+}1) \gets \min(\max(\mathbf{x}^{(p)}(t{+}1),0),1)$ 
        
        $\mathbf{b}^{(p)}(t{+}1) \leftarrow \text{Top}_{N_{\mathrm{T}}-|\mathcal{F}|}(\mathbf{x}^{(p)}(t{+}1)) \cup \mathcal{F}$
        
        $f^{(p)}(t{+}1) = f(\mathbf{b}^{(p)}(t{+}1))$ \tcp{ Refer Eq. \eqref{eqn:PSLL}}
        \If{$f^{(p)}(t{+}1) < f(\mathbf{P}_{\mathrm{best}}^{(p)})$}{
            $\mathbf{P}_{\mathrm{best}}^{(p)} \gets \mathbf{x}^{(p)}(t{+}1)$
        } 
        \If{$f^{(p)}(t{+}1) < f(\mathbf{b}(\mathbf{G}_{\mathrm{best}}))$}{
            $\mathbf{G}_{\mathrm{best}} \gets \mathbf{x}^{(p)}(t{+}1)$ 
        }
    }
}
$\mathbf{b}_{\mathrm{opt}} = \text{Top}_{N_{\mathrm{T}}-|\mathcal{F}|}(\mathbf{G}_{\mathrm{best}}) \cup \mathcal{F}$%
\end{algorithm}
\subsection{PSO-based Array Thinning for Sum-Rate Maximization}
In this subsection, we explain the \ac{STA} to maximize the achievable sum-rate in \eqref{eqn_IIB_1}. We assume that range and angle information of the \acp{UE} is known at the \ac{BS}. For $K$ \acp{UE} with $\boldsymbol{\theta}=[\theta_{1},\ldots,\theta_{K}]$ and ranges $\mathbf{r}=[r_{1},\ldots,r_{K}]$, we construct the channel matrix $\mathbf{H}\in\mathbb{C}^{N\times K}$ based on \eqref{eqn_IIA_2}. Similar to the structure of the grating-lobe problem, we enforce a fixed number of active $N_\mathrm{T}$ antennas and a set $\mathcal{F}$ of mandatory active indices. The optimization problem \eqref{eq:gen_opt_problem} is reformulated as sum-rate maximization problem, where $f(\mathbf{b})=\mathcal{R}_{\mathrm{sum}}(\mathbf{b})$, and is given by 
\begin{equation}\small
    \max_{\mathbf{b}}
    \quad
    \mathcal{R}_{\mathrm{sum}}(\mathbf{b}),
    \text{s.t.}\quad 
    \sum_{n=1}^{N} b_{n} = N_{\mathrm{T}},
     \quad
    b_{n} = 1\quad n\in\mathcal{F}.
    \label{opt:sumrate-max} 
\end{equation}
\normalsize

\begin{figure}[t]
\centering
    \includegraphics[width=1\linewidth]{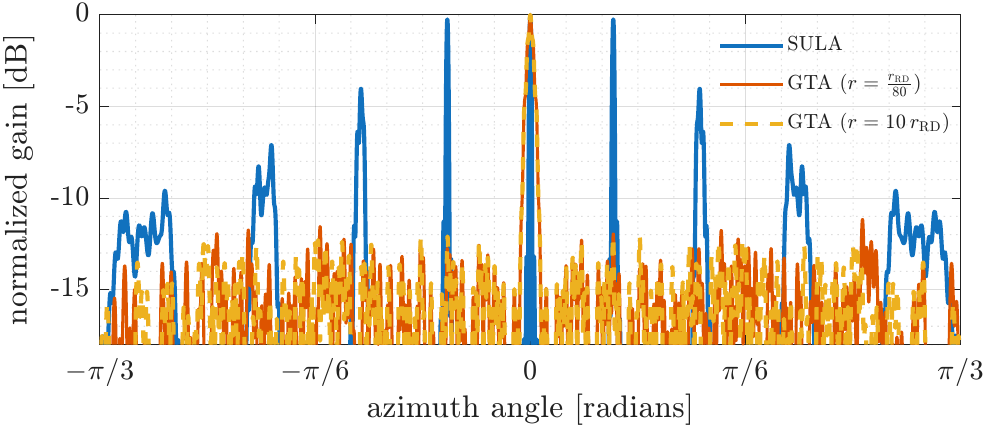}
\setlength{\belowcaptionskip}{-15pt}
\setlength{\abovecaptionskip}{-10pt}
\caption{Grating lobe suppression in the angle domain. The mainlobe is at $\theta=0^\circ$ and $\RD=\unit[508]{m}$. }
\label{fig:GL_suppresion}
\end{figure}

The optimization problem in \eqref{opt:sumrate-max} is combinatorial and NP-hard. Therefore, we adopt the same continuous-relaxation PSO framework described in the previous subsection. Each particle $p$ maintains a continuous priority vector $\mathbf{x}^{(p)}\in[0,1]^{N_{v}}$, where $N_{v}=N-|\mathcal{F}|$ denotes the number of variable antenna positions. A binary thinning vector $\mathbf{b}^{(p)}$ is obtained by activating all fixed indices and selecting $(N_{\mathrm{T}}-|\mathcal{F}|)$ entries with the largest values in $\mathbf{x}^{(p)}$. The main distinction from the grating-lobe suppression formulation lies in the objective: the cost function is now the sum-rate ($f(\mathbf{b})=\mathcal{R}_{\mathrm{sum}}(\mathbf{b})$), whereas all PSO update rules remain unchanged. In particular, in line 10 of Algorithm~\ref{alg:PSO_thinning}, the cost is computed using the sum-rate expression in~\eqref{eqn_IIB_1}.

\begin{figure*}[!t]
  \centering
  \begin{minipage}[b]{0.31\textwidth}
    \centering
    \includegraphics[width=\textwidth]{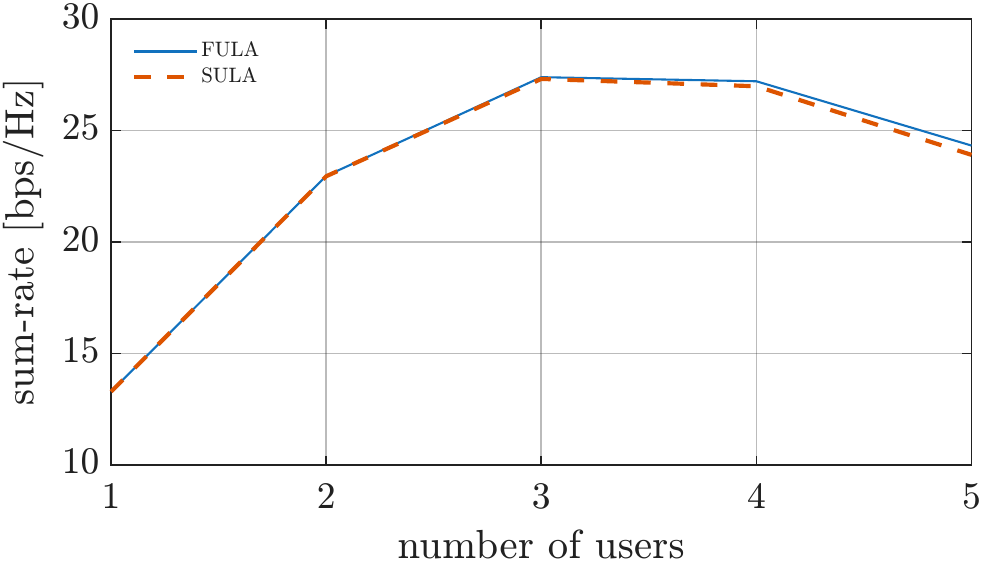}
\caption{Sum-rate for \ac{SULA} when \acp{UE} are distributed only along the range.}
    \label{fig:SumRate_vs_range}
  \end{minipage}%
  \hspace{2mm}
  \begin{minipage}[b]{0.31\textwidth}
    \centering
    \includegraphics[width=\textwidth]{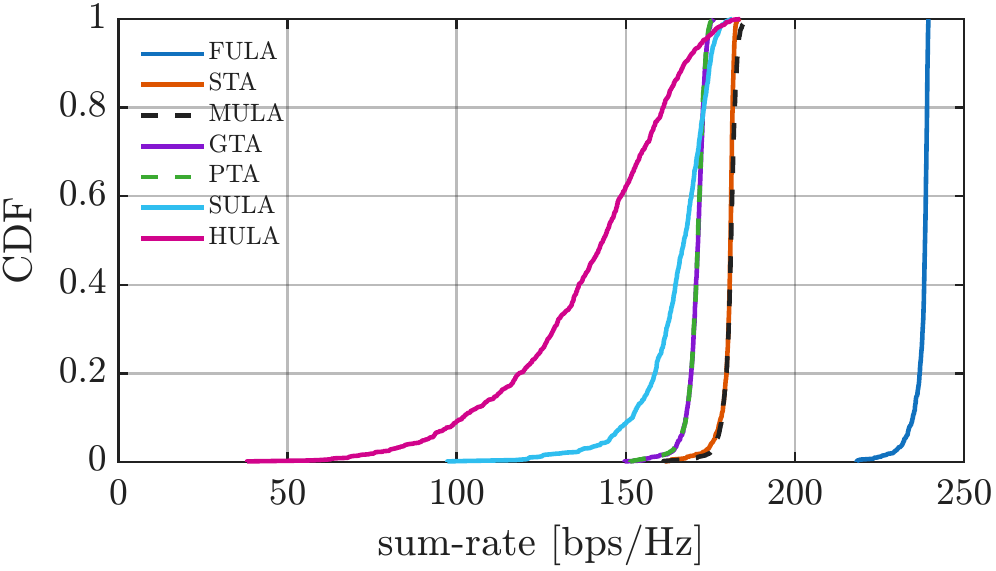}
\caption{Cumulative distribution function of the sum-rate across different sparse arrays.}
    \label{fig:sumrate_vs_CDF}
  \end{minipage}
  \hspace{2mm}
  \begin{minipage}[b]{0.31\textwidth}
    \centering
    \includegraphics[width=\textwidth]{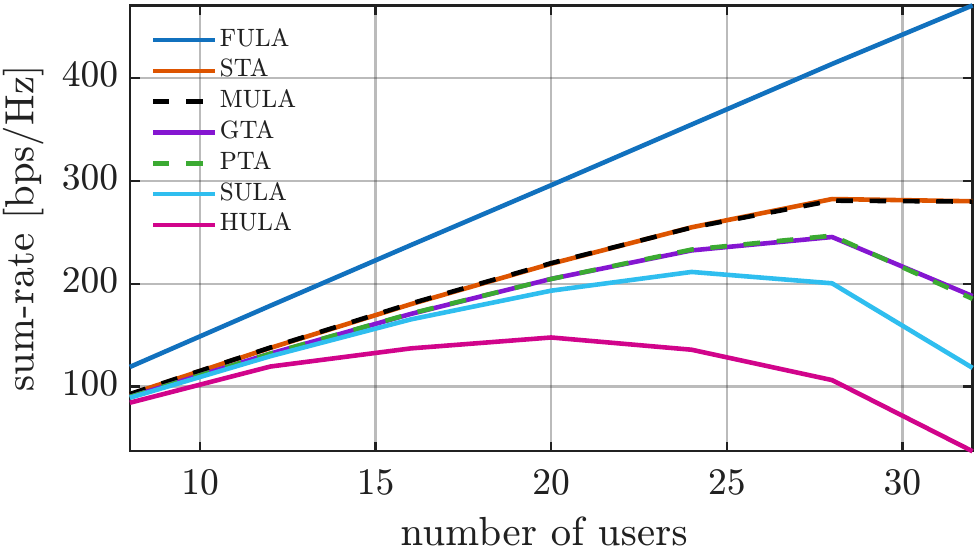}
    \caption{Average sum-rate vs. number of users for different sparse arrays.}
    \label{fig:sumrate_numofusers}
  \end{minipage}
  \vspace{2mm}
  \vspace{-0.35in}
\end{figure*}

\section{Simulation Results} \label{Sec-V}
We evaluate the performance of the proposed \ac{GTA} and \ac{STA} designs by comparing them against several benchmark array configurations. In our setup, the \ac{BS} employs a \ac{FULA} with $N=320$ antennas, while the thinned configurations activate only $N_{\mathrm{T}}=32$ antennas, yielding a thinning ratio of $\mathrm{TR}=\tfrac{1}{10}$. In general, there is no single optimal number of active antenna elements, as it depends on system requirements. Increasing the number of active elements improves the beamforming gain and brings the sparse array performance closer to that of the \ac{FULA}, at the cost of higher hardware complexity and power consumption.
The carrier frequency is $\unit[30]{GHz}$ and \ac{SNR} is set to $\unit[20]{dB}$. The following benchmark arrays are considered:
\begin{itemize}
    \item \textbf{\ac{FULA}:} A conventional ULA with $N=320$ antennas and spacing $d=\tfrac{\lambda}{2}$, serving as an upper bound.
    \item \textbf{\ac{MULA}:} A movable ULA with $N=32$ antennas where antenna positions are optimized within $[-80 \lambda, 80\lambda]$ using \ac{PSO} for each channel realization.
    \item \textbf{\ac{PTA}:} A pre-optimized thinned array with $N_{\mathrm{T}}=32$ active antennas selected using statistical \ac{CSI}, following the approach presented in \cite{11030818}.
    \item \textbf{\ac{SULA}:} A sparse ULA with $N_{\mathrm{T}}=32$ antennas and uniform spacing $d = 5\lambda$, chosen to match the aperture of the \ac{FULA}.
    \item \textbf{\Ac{HULA}:} A compact ULA with $N_{\mathrm{T}}=32$ antennas and $d=\tfrac{\lambda}{2}$.
\end{itemize}
Except for the compact \ac{HULA}, all configurations share the same aperture length to ensure a fair comparison. We keep the \ac{PSO} parameters as given in \cite{11030818}. First, we illustrate the grating lobe suppression performance of \ac{GTA} with $N_T=32$, where the mainlobe is focused at boresight. Fig.~\ref{fig:GL_suppresion} shows that \ac{SULA} exhibits strong grating lobes, while \ac{GTA} effectively suppresses them at both $\frac{\RD}{80}$ and beyond $\RD$, demonstrating that the proposed approach is effective across all ranges. Next, we consider with a scenario where all \acp{UE} are aligned at the same boresight angle $\theta=0^\circ$ and randomly distributed along the range axis, i.e., $r \sim \mathcal{U}[2D = \unit[3.18]{m},\, \tfrac{\RD}{7} = \unit[72.6]{m}]$, where $\tfrac{\RD}{7}$ is the maximum beamfocusing distance at boresight \cite{10988573}. Fig.~\ref{fig:SumRate_vs_range} compares the sum-rate performance of the \ac{FULA} and the \ac{SULA}, with the beamforming gain normalized by the number of antenna elements. The two configurations exhibit nearly identical performance because (i) they share the same physical aperture and (ii) grating lobes do not occur along the range dimension for the \ac{SULA}. Furthermore, since the \acp{UE} are aligned in angle but separated in range, the presence of grating lobes in the angular domain does not affect the sum-rate performance. Next, we consider downlink transmission to $K=16$ \acp{UE}, whose polar coordinates are generated according to $r \sim \mathcal{U}[2D = \unit[3.18]{m},\, \tfrac{\RD}{7} = \unit[72.6]{m}]$ and $\theta \sim \mathcal{U}[-\tfrac{\pi}{3},\, \tfrac{\pi}{3}]$. Fig.~\ref{fig:sumrate_vs_CDF} illustrates the \ac{CDF} of the achievable sum-rate for all considered schemes. As expected, the \ac{FULA} delivers the highest performance due to its full aperture and maximum beamforming gain. The proposed \ac{STA} achieves performance comparable to that of the \ac{MULA}. Moreover, it attains approximately $75\%$ of the \ac{FULA} sum-rate while utilizing only $10\%$ of the active elements. The \ac{GTA} attains performance comparable to the \ac{PTA}, indicating that grating-lobe suppression effectively reduces interference and enhances sum-rate. Although \ac{GTA} removes grating lobes, its sidelobes remain relatively elevated due to irregular element spacing and the reduced number of active antennas. The sum-rate of \ac{GTA} and \ac{PTA} is on average $5\%$ lower than that of the \ac{STA}. However, both the \ac{GTA} and \ac{PTA} are pre-optimized, whereas the \ac{STA} requires more frequent updates depending on the channel geometrical parameters. Finally, Fig.~\ref{fig:sumrate_numofusers} shows the average sum-rate versus the number of served \acp{UE}. The proposed \ac{STA} consistently outperforms all sparse baselines. For moderate system loading, i.e., when $\tfrac{N_{\mathrm{T}}}{K} > 1$, the sum-rate increases approximately linearly and gradually saturates as $\tfrac{N_{\mathrm{T}}}{K}$ approaches unity, due to the transition to an interference-limited regime. Furthermore, the proposed \ac{STA} achieves sum-rate performance comparable to the \ac{MULA}. Importantly, array thinning provides a hardware-efficient solution and enables graceful performance degradation, since faulty elements can be deactivated and the thinned array configuration can be dynamically re-optimized. 

The computational complexity of the \ac{PSO} algorithm in Algorithm~\ref{alg:PSO_thinning} scales with the number of particles $P$, the number of iterations $n_{\mathrm{PSO}}$, and the number of optimization variables $N_v$, resulting in an overall complexity of $\mathcal{O}(n_{\mathrm{PSO}} P N_v)$. The proposed \ac{STA} is more computationally efficient than \ac{MULA}, since the latter requires a larger number of iterations to converge due to its continuous search space. Specifically, $n_{\mathrm{PSO}} = 100$ for \ac{STA} and $n_{\mathrm{PSO}} = 200$ for \ac{MULA}. Furthermore, computational complexity is less critical for \ac{PTA} and \ac{GTA}, as their configurations are optimized offline before deployment. 

\section{Conclusion}\label{Sec-VI}
This work introduced a dynamic array thinning framework to optimize multi-user sum-rate in the near field. The proposed \ac{STA} achieves performance comparable to the \ac{MULA} while avoiding the hardware challenges associated with \ac{MULA}. Future work will explore replacing the \ac{PSO} with a deep learning-based solution to mitigate the computational complexity of the current optimization approach.

\bibliographystyle{IEEEtran}
\bibliography{IEEEabrv,my2bib}
\end{document}